\documentstyle[12pt,epsfig]{article}

\def\Title#1{\begin{center} {\Large {\bf #1} } \end{center}}

\begin{document}

\Title{High Energy Particles from the Universe}

\bigskip\bigskip


\begin{raggedright}  

{\it Rene A. Ong\index{Ong, R.A.}\\
Enrico Fermi Institute \\
University of Chicago,
Chicago, IL 60637 }
\bigskip\bigskip
\end{raggedright}

\section{Outline}

It is an exciting time to be working at the interface between
physics and astronomy.
Experiments built to detect astrophysical and atmospheric
sources of neutrinos, such as 
Super-Kamiokande\index{Super-Kamiokande},
are providing what may be the first definitive evidence for
neutrino mass.
Cosmology experiments, such as those measuring the
the redshifts of type Ia supernovae and those determining the
anisotropy of the microwave background,
may have an important impact on particle physics as they
confront our general view of
the origin of the Universe.
There have been similarly exciting developments at the high energy
frontier of astronomy.
Among other things:
\begin{enumerate}
\item we have detected
a gamma-ray burst that is probably the most
powerful explosion recorded since the Big Bang,
\item we have discovered extragalactic
astrophysical sources that beam intense fluxes of 
TeV $\gamma$-radiation to us, and
\item we have observed 
individual particles (possibly protons) 
arriving from outer space
with energies exceeding 25 Joules.
\end{enumerate}

\noindent These new developments are clearly 
of prime importance for the
field of high energy astrophysics, but, even more,
research in this area in the future
may make important contributions to
particle physics depending, of course, on the nature of
the discoveries.

In this paper, I provide a summary of the field
of high-energy astronomy using photons, cosmic rays, and neutrinos.
This field is rapidly developing because of recent discoveries and
because of new experimental techniques that derive largely from
accelerator-based particle physics detectors.
I start with
a broad overview of the field and discuss the
general scientific motivations.
Then, the
three exciting developments listed above are
described in more detail, followed by a
discussion of the theoretical considerations.
A selective review of the field in terms of the experimental techniques,
results, and future prospects
makes up the last part
of the paper.
I conclude with a summary of the prospects for the future.

\section{Broad Overview}

We learn about the Universe outside the
immediate neighborhood of the Solar System by studying the arrival of
four distinct messengers:
1) photons,
2) cosmic rays,
3) neutrinos, and
4) gravity waves \cite{Harwit}.
Here, cosmic rays\index{cosmic rays} are defined
as nuclei (p, n, He$^{++}$, etc.)
electrons, and their antiparticles.
In the future, we may discover other stable particles
that convey information across interstellar
space. 
If we do, we can add them to the previous list.

To date, the detection the photons over a wide range of energies
has been the basis for the vast majority of astronomical discoveries.
Cosmic rays provide important information about high energy processes
occurring in our galaxy.
Neutrinos and gravity waves both offer great astronomical potential,
but are difficult to detect.
So far, the neutrino source list is limited to our Sun and the
supernova SN1987A.

Photon, cosmic ray, and neutrino astronomy are closely related
at high energies (E$ >1\,$GeV), both in terms of their astrophysical production
mechanisms and in terms of their detection techniques.
High energy particles are produced astrophysically
by acceleration 
processes rather than by thermal processes which dominate
at lower energies.
The high energy particle fluxes typically exhibit rapidly falling
power-law spectra, which leads to
the requirement of very large detectors.
The high energy domain spans a wide dynamic range
(from 1\,GeV to $10^{11}$\,GeV), and we cannot expect a single
detection technique to work at all energies.

In the future, we imagine studying astrophysical sources with multiple
messengers that provide complementary information.
A prototypical source would be gamma-ray bursts (GRBs).
We know a great deal about GRBs from the electromagnetic radiation
they produce, but there is also speculation that they are the source
of the highest energy cosmic rays \cite{Waxman} and that
they produce a detectable neutrino flux
\cite{Halzen1}.
Thus, we anticipate that astronomy using these different
messengers will become interrelated as the experiments become more
powerful and as more detections are made.



The general scientific motivation for 
particle astronomy is multi-faceted.
On the physical side, we use high energy
radiation to probe extreme conditions of magnetic 
or gravitational potential.
Our understanding of such
astrophysical situations is still far from complete.
The copious flux of
high energy cosmic rays argues for efficient
astrophysical accelerators with beam energies well beyond what
we can achieve on Earth.
High energy particle astronomy may also shed light on aspects of
particle physics or cosmology beyond their respective standard models. 
On the astronomical side, the movement towards high energies
continues the historical expansion of astronomy from optical light
into new wavebands (radio, infrared, X-ray, and $\gamma$-ray).
It is important to emphasize, however, that
when discussing
the many aspects of high energy astronomy, there
is not yet a generally applicable ``Standard Model''.
For the experimentalist, this situation is ideal in that
there are few constraints.
New experimental results often change the general paradigm.

\section{Three Recent Exciting Results}

Here I discuss three of the most exciting results in this
field in the last few years.
The choices are, of course, very subjective.

\subsection{Gamma Ray Bursts}
Even after thirty years of research,
the nature of gamma ray bursts (GRBs) \index{gamma ray bursts}
 remains one
of the most important mysteries of astrophysics.
We know from the BATSE \index{BATSE}
detector on the Compton Gamma Ray
Observatory that the arrival directions of GRBs
are consistent with isotropy \cite{Fishman}.
However, the relatively poor angular resolution
of BATSE of a few degrees has made the
detection of GRB counterparts
(i.e. sources at optical or radio
wavelengths)
difficult.
The lack of counterparts has hindered efforts to pin down
the distance scale of GRBs.

In 1997, a major breakthrough was achieved with the first detection
of optical counterparts.
This work was accomplished by the Beppo/SAX \index{Beppo/SAX}
satellite detector
in conjunction with powerful optical telescopes, both
on the ground (e.g. Keck) and in space (e.g. HST).
By using a combination of wide and narrow-field X-ray telescopes,
Beppo/SAX determines the positions of some bursts with an
accuracy of several arc minutes.
This excellent localization allowed
the detection of fading optical counterparts 
for several dozen bursts \cite{Hurley}.
Redshift values have been determined for approximately
twenty counterparts; they indicate that the correlated bursts
are cosmological in origin (typical $z \sim 1$).
The typical inferred energy outputs of the bursts 
range between $10^{51}$ and $10^{53}$ ergs
(assuming isotropic emission).

The detection of optical counterparts to GRBs is clearly
a landmark discovery.
Until 1999, however,
optical observations were limited to time periods
well after the actual bursts because of the long
delay (typically 8 hours) required by Beppo/SAX
to achieve accurate position localization.
What we really want to do is to carry out
out {\em simultaneous} $\gamma$-ray and optical
observations of GRBs.
Such observations require a fast slewing telescope with
a wide field of view to cover the large BATSE error box.

Several wide-field optical telescopes have been constructed or
are currently under development.
One of the pioneering experiments, known as ROTSE-I,
consists of an array of 35\,mm telephoto lenses coupled
to large format CCD detectors.
With an overall field of view of $16^\circ\,$x$\,16^\circ$, and
by slewing in an automated fashion upon receiving an alert
from the GRB Coordinates Network (GCN),
ROTSE \index{ROTSE}
is ideally suited for rapid optical follow-up
of GRBs.
The careful design of the experiment paid off on January 23, 1999
when ROTSE made the first detection of contemporaneous optical
radiation from a GRB \cite{Akerlof}.

\begin{figure}[htb]
\begin{center}
\epsfig{file=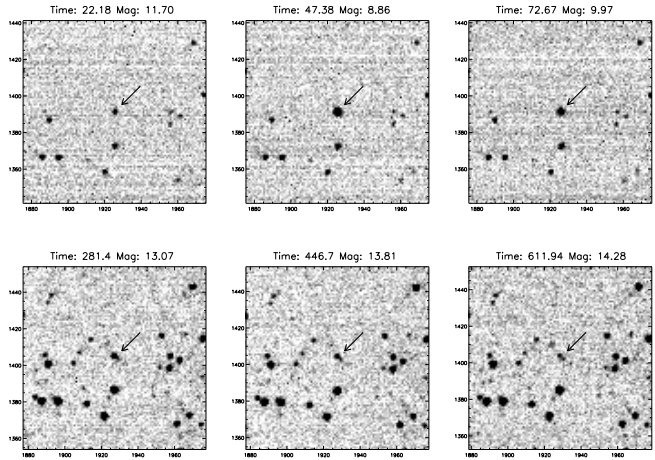,height=3.5in,angle=90}
\caption{
Detection of powerful optical radiation from 
a gamma ray burst in progress
\protect\cite{Akerlof}.
Six successive CCD images taken by the ROTSE-I
experiment are shown for GRB990123.
At a redshift of 1.60 and a peak magnitude of
8.95, this burst was the most luminous astronomical
object ever detected.}
\end{center}
\label{ROTSE}
\end{figure}

From the perspective of BATSE, 
GRB990123 was relatively ordinary,
but the optical signature detected by ROTSE, 
was truly remarkable.
As shown in Figure~\ref{ROTSE},
the optical brightness increased by 3 magnitudes
in 25 seconds and then waned by 5 magnitudes over a period of
8 minutes.
At its peak brightness, the optical magnitude was 8.95
which, when combined with the measured redshift
value of 1.60, meant that this burst was the most
luminous object ever detected.
GRB990123 was brighter than the brightest quasar
by several orders of magnitude.
Assuming isotropic emission,
the inferred energy release of GRB990123 exceeds
$10^{54}\,$ergs.
Theoretical models, already grappling to explain the
wide variety of GRB phenomena, are further strained to deal
with this remarkable energy output.

A wide range of theoretical models have been proposed to explain
gamma ray bursts.
There are almost as many models as bursts!
The basic difficulty is to construct a physical mechanism that can
produce and extract the intense high energy emission we observe.
The general picture calls for a cataclysmic event which produces
a relativistic fireball \index{fireball model of gamma ray bursts}
of material with Lorentz factors approaching
1000 \cite{Meszaros}.
The relativistic material escapes from the region of high 
energy density along a collimated jet.
High energy radiation results when the jet collides
with nearby ambient material.
The nature of the original cataclysmic event is not fully understood.
Generally favored pictures include colliding neutron stars
and hypernovae (``failed supernovae'' of heavy stars) \cite{Woosley}.

The recent discoveries in the area of gamma ray bursts are clearly
profound, but the overall puzzle is far from being solved.
Many additional questions remain.
For example,  when considering the 
time durations and spectral shapes of GRBs,
we know from the BATSE data that there are
at least two classes of bursts, (if not more)
\cite{Nemiroff}.
For instrumental reasons, Beppo/SAX
is sensitive only to longer bursts (duration greater
than 1 second).
Thus, so far, we can confidently ascribe a cosmological
origin only to a portion of the burst population.
The medium and short length bursts may have a different
origin.
We can expect GRB research to remain exciting
for years to come.

\subsection{TeV $\gamma$-rays from Extragalactic Sources}

The field of very high energy (VHE) $\gamma$-ray astronomy
has come of age in the last ten years \cite{Ong,Weekes}.
Ground-based telescopes using the atmospheric Cherenkov
technique have made strong detections of TeV photons from 
galactic and extragalactic sources.
The most exciting development in this area has been the
discovery of strong VHE emission from active galactic nuclei (AGN),
which make up a broad class of extragalactic objects including quasars.
AGN \index{AGN-active galactic nuclei}
are powerful sources at all wavelengths, but the importance of
their $\gamma$-ray output has only recently been fully appreciated.
We now know that in some AGN, those known as {\em blazars},
the bulk of their power is emitted
at high energies.
\index{blazars}
Explaining this observational fact is an important problem facing
theorists.

\begin{figure}[htb]
\begin{center}
\epsfig{file=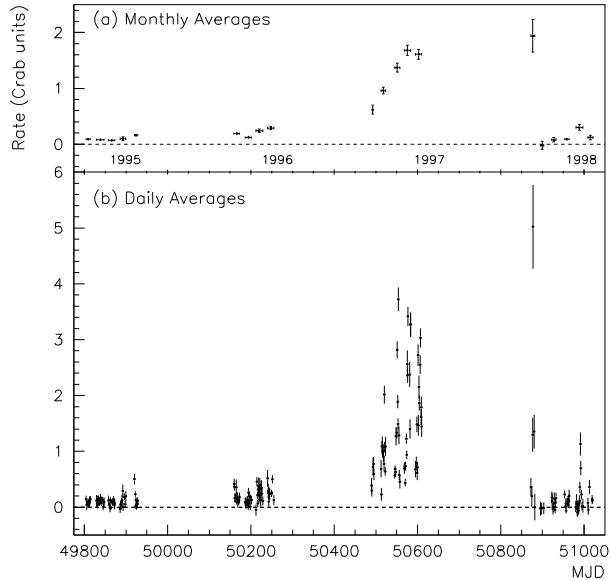,height=3.0in}
\caption{
The flux of very high energy $\gamma$-rays from the
extragalactic blazar source Markarian 501
\protect\cite{Quinn}.
The flux over a four year period as
recorded by the Whipple Observatory is shown
normalized by the flux from the Crab Nebula
(the brightest known galactic source).
In 1997,
dramatic increases in the flux level and in the
degree of variability were recorded.
The $\gamma$-ray energies range between $250\,$ GeV and
$20\,$TeV.}
\end{center}
\label{mrk501}
\end{figure}

In 1997, the blazar source Markarian 501 (Mrk 501)
entered a period of dramatic
activity.
\index{Markarian 501}
The TeV $\gamma$-ray emission from the source increased by up
to a factor of fifty from earlier epochs.
Figure~\ref{mrk501} shows the VHE $\gamma$-ray flux detected from this source
over a four year period by the Whipple experiment.
The difference between the average flux level seen in 1997 and in the other
years is striking.
At certain times, the $\gamma$-ray flux from Mrk 501 exceeded the
brightest known source in our galaxy, the Crab Nebula, by 
factors of three to four.
\index{Crab Nebula}
This happened in spite of the fact that 
Mrk 501 is more distant from us than
the Crab by a factor of 10,000.
At its maximum brightness, Mrk 501 beamed 10$^{11}$ VHE photons
per second to the Earth's surface.
The VHE spectrum extends to maximum
energies above $20\,$TeV.

Another impressive feature of the 1997 emission from Mrk 501 was
the high degree of variability observed.
The $\gamma$-ray flux varied by as much as an order of magnitude
from night to night and by factors of two on hourly time scales.
These variations imply that the acceleration region where the
$\gamma$-rays are produced must be very compact, 
presumably several light
hours across (times any relativistic Doppler factors).
The flux variations of Mrk 501 were studied by several state-of-the-art
Cherenkov telescopes around the globe and there was good agreement between
the flux levels recorded by the different experiments.

\begin{figure}[htb]
\begin{center}
\epsfig{file=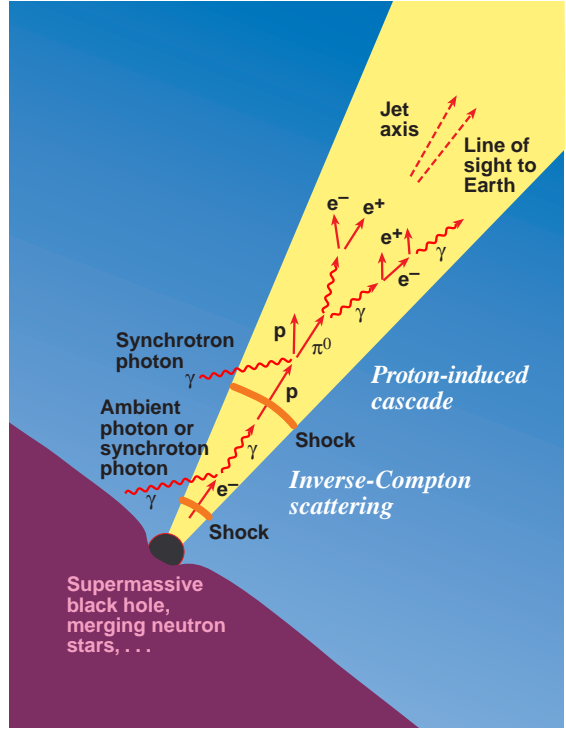,height=3.8in}
\caption{Artist's conception
of the acceleration processes in a blazar
\protect\cite{Halzen2}.
Black hole accretion powers relativistic jets of material.
In the jets, electrons or protons are accelerated to high
energies via shocks.
High energy $\gamma$-rays are produced from the inverse-Compton
scattering of electrons or from cascades initiated
by protons.
We detect emission from those sources 
which are beamed into our line of sight.}
\end{center}
\label{halzen}
\end{figure}

AGN are such luminous objects that the only feasible source
for their power is in the intense gravitational
field near black holes.
The general picture of an AGN is a supermassive
($10^8 - 10^9$\,M$_\odot$) black hole surrounded by
a rotating disk.
Accreting material provides the power for the broad-band
emission observed.
It also powers relativistic
jets directed along the angular momentum axis of the black hole.
As shown in Figure~\ref{halzen},
shock acceleration in the jets produces high energy beams of
electrons or protons which interact with radiation fields to
produce secondary beams of
$\gamma$-rays and neutrinos.
A key aspect of this model is that
we observe blazars when their beams are directed into our
line of sight.

In the general blazar picture, 
the high energy $\gamma$-rays probe acceleration processes 
in the jet.
The fact that very rapid $\gamma$-ray variability has been observed 
indicates that the acceleration region 
relevant to high energy particle production
is deep within the jet, and possibly quite
close to the black hole itself.
Thus, understanding the VHE $\gamma$-ray emission is of
crucial interest.

\subsection{Cosmic Rays with Energies Exceeding $10^{20}\,$eV}

We have known about the existence of extremely high energy
($E > 10^{20}\,$eV)
cosmic rays for more than forty years.
In 1966, it was realized by
Greisen, Zatsepin, and Kuz'min (GZK)
\cite{GZK}
\index{GZK cutoff}
that cosmic rays above the energy of $\sim 6\times 10^{19}\,$eV
(if they existed!)
would interact with the 3\,K cosmic microwave
background radiation (CMBR).
This GZK cutoff
limits the mean free path for
the highest energy cosmic rays to less than 100 Mpc, a distance that
is quite small in comparison to typical extragalactic scales.
Until recently, conventional wisdom held
that cosmic rays above the GZK cutoff would not be detected on Earth
because: 1) it was hard to construct astrophysical sources
capable of accelerating particles to these energies,
and 2) if such source existed, they would almost certainly
be located at great distances and the cosmic rays they produced
would be absorbed by the CMBR.
This wisdom proved to be wrong.  
It now seems clear that
the cosmic ray spectrum continues 
to energies of $10^{20}\,$eV,
and beyond.

The most compelling evidence for particles beyond the GZK cutoff
comes from the AGASA experiment \cite{AGASA1}.
\index{AGASA}
AGASA is a large surface array covering an area of approximately
$100\,$km$^2$, and
located in the central portion of the
Japanese island of Honshu.
The array samples the particle cascade in giant air showers 
produced from extremely high energy cosmic ray interactions
in the atmosphere.
The cosmic ray energy is estimated from the particle density
determined a fixed distance from the core of the shower.
Uncertainties in the energy measurement are estimated by
several techniques, including simulation methods.
The typical energy resolution is $\sim 30$\%. 
More importantly, the proportion
of events with a 50\% or more overestimation in energy
is less than 3\%.

\begin{figure}[htb]
\begin{center}
\epsfig{file=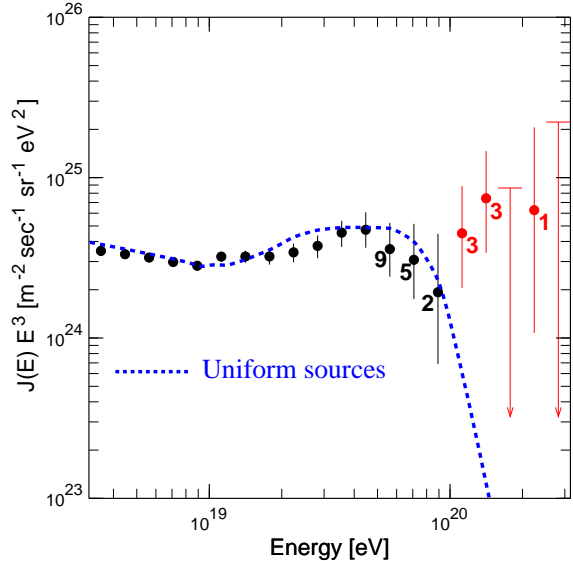,height=2.8in}
\caption{Energy spectrum of the highest energy cosmic rays
measured by the AGASA experiment \protect\cite{AGASA1}.
As indicated, seven events are detected above
$10^{20}\,$eV.
The dashed line corresponds to the spectrum expected from
a population of extragalactic sources distributed uniformly
in the Universe.
For display purposes,
the differential particle flux is multiplied by the energy cubed.}
\end{center}
\label{agasa}
\end{figure}

The latest cosmic ray energy spectrum released by
AGASA is shown in Figure~\ref{agasa}
\cite{AGASA1}.
Based on a data sample collected between 1993 and 1998,
the spectrum shows no evidence for a cutoff above
$6\times 10^{19}\,$eV.
Instead, seven events are detected above 
$10^{20}\,$eV.
The measured spectrum from AGASA is inconsistent with
that expected from a population of extragalactic sources
distributed uniformly in the Universe.
Barring an unforeseen loophole to the GZK cutoff,
the AGASA results, in conjunction with the detection by the
Fly's Eye experiment \index{Fly's Eye}
of a cosmic ray at $3\times 10^{20}\,$eV
\cite{FlysEye}, strongly support the idea the sources
of the highest energy cosmic ray events are ``local''.

Considering the enormous rigidity of $10^{20}\,$eV
particles and the weakness of the intergalactic magnetic field
($B < 10^{-10}\,$g),
we do not expect substantial deflection of such particles
over path lengths of 50\,Mpc.
Thus, it is reasonable to consider doing astronomy with
the extremely high energy cosmic rays.
The arrival directions of the cosmic rays above
$10^{19}\,$eV have been examined. 
No obvious concentration
in the sky can be discerned.
Also,
there is no significant correlation between the cosmic ray positions
and those of known astronomical sources (e.g. quasars,
radio galaxies, etc.).
Thus, we are faced with a quandary similar to that of gamma ray bursts.
Since the sources of the highest energy cosmic rays cannot be
correlated with known astronomical objects, they must represent something
new.
To achieve their extreme energies, the cosmic rays must be
produced in a truly remarkable astrophysical accelerator.
They could, in fact, come from physics at a higher mass scale.
It is perhaps this latter possibility that makes these particles so
intriguing.

\section{Theoretical Considerations}

\subsection{Extreme Astrophysics}

As introduced in Section~2,
we expect high energy $\gamma$-rays, cosmic rays,
and neutrinos to be produced 
via particle interactions at sites of powerful acceleration.
When discussing such production, we must distinguish between
the {\em power source} and the 
{\em acceleration mechanism}.
For sources using conventional (i.e. known) physics,
the power source will make use of extreme 
electromagnetic or gravitational potentials.
Pulsars, such as the Crab, \index{Crab Nebula}
are examples of the former case.
Supernova remnants \index{supernova remnants}
and AGN are examples of the latter.

High energy $\gamma$-radiation has been detected from a number
of known radio pulsars.
The radiation can be categorized as being {\em pulsed} 
(i.e. with the same period as that detected in the radio)
or {\em unpulsed}.
For both categories, the ultimate power source
derives from the spin-down of a highly magnetized neutron star.
At the highest energies,
the radiation is believed to originate from 
inverse-Compton scattering of soft photons by a relativistic
wind of electrons \cite{Harding2}.
In the case of the Crab, 
from which $\gamma$-rays up to 50\,TeV
have been detected, the electron spectrum is believed to extend above
$1\,$PeV, making the Crab
the high-energy accelerator known in the Universe.

Supernova remnants (SNRs) are attractive candidates for
particle acceleration because of the
enormous power contained in the original explosion.
Supernova explosions typically release $10^{51}\,$ergs
of kinetic energy and they occur every few decades
somewhere in the galaxy.
Therefore, the average power injected into the interstellar
medium by these explosions is $\sim 10^{42}\,$ergs\,s$^{-1}$.
Since the power required to replenish the cosmic rays
is $10^{40} - 10^{41}\,$ergs\,s$^{-1}$, 
supernova remnants can explain the origin of the cosmic rays if
$1 - 10$\% of the kinetic energy released in the explosion
goes into the
acceleration of protons and nuclei.

Particle acceleration in SNRs
is expected to occur at the shock front produced 
as the remnant material traverses the interstellar medium.
Charged particles scatter off magnetic field irregularities
as they diffuse back and forth across the shock front.
Because the velocity distributions of the scattering centers
on either side of the shock are isotropic, the particles
see a converging flow of scattering centers on
both sides of the front.
They thus gain energy
with each round trip passage.
This particle acceleration process, first proposed by
Fermi in a different context \cite{Fermi},
continues as long as the particles are contained in the
vicinity of the shock front.
The containment time is proportional to energy, and so
diffusive shock acceleration naturally leads to a power
law energy spectrum for the accelerated particles,
(dN/dE\,$\propto\,$E$^\alpha$).
For strong (highly supersonic) shocks, the power law
spectral index is $\alpha \sim -2$.

The maximum obtainable particle energy in SNRs
is determined by the lifetime of the shock itself.
As the shock wave expands, it slows downs and weakens.
For typical shock lifetimes ($\sim 5000\,$yr), the
maximum particle energy is calculated to be
Z$\times 10^{14}\,$eV, where Z is the particle charge
\cite{Legage}.
Thus, SNRs offer a
plausible explanation for the origin of cosmic rays
up to an energy of $10^{15}\,$eV (and possibly up to
$10^{16}\,$eV),
but a new source is 
required to explain higher energy
cosmic rays.

There is a great deal of speculation about the
origin of particles above $10^{16}\,$eV.
Galactic objects do not in general have the
combination of size and magnetic field strength
to contain a particle at these energies
\cite{Hillas}.
Therefore, it is generally believed that the
sources must be extragalactic.
AGN are attractive possibilities because they
are known to be powerful emitters of $\gamma$-rays.
They may also produce very high energy cosmic rays
and neutrinos.

The general model for AGN
was outlined in Section~3.2
(see Figure~\ref{halzen}).
Charged particles are accelerated to high energies
in the jet via the Fermi shock mechanism.
The accelerated particles are likely to be
electrons or protons (or a combination of the two).
Electrons are certainly needed to explain the broad-band
synchrotron emission observed.
If electrons dominate,
$\gamma$-rays could naturally be produced by
inverse-Compton scattering off low energy
photons \cite{Maraschi}.
The soft photons may originate from synchrotron radiation,
from disk emission, or from reprocessed emission from clouds
or dust \cite{Dermer}.
If protons dominate, they would interact with radiation fields to
produce electromagnetic cascades that would ultimately
produce high energy $\gamma$-rays \cite{Mannheim}.
Protons may also produce neutrinos via a 
beam dump mechanism
(i.e. by impinging on
material to produce charged pions which then decay).


Extragalactic sources, such as AGN, may explain the
origin of cosmic rays between $10^{16}\,$eV and $10^{19}\,$eV.
However, as discussed in
Section~3.3, the highest energy particles 
cannot come from large extragalactic distances.
Their origin is a true mystery.
Given our inability to explain their origin
via known astrophysics,
we are forced to seriously consider other
possibilities, such as new astrophysics or particle physics
beyond the Standard Model.

\subsection{Non-Standard Astrophysics or Particle Physics}

When explaining the existence of particles arriving at Earth with
very high energies, we must
naturally consider possible sources outside known astrophysics.
In this case, particles would acquire their energies
in a ``top down'' picture,i.e. as coming from physics at
a higher mass scale, as opposed to the ``bottom up'' picture
of acceleration.
Cosmology must come into play here, because the early Universe
offers the most natural conditions needed to generate
interactions or particles beyond the Standard Model.
Here we give a few examples of non-standard sources of very high
energy particles.

There is speculation that the highest energy
cosmic rays result from
the collapse of topological defects produced in the early Universe
\cite{Schramm}.
\index{topological defects}
Defects such as cosmic strings resulting
from phase transitions could
\index{cosmic strings}
produce grand
unified (GUT) scale particles at with masses of
$10^{14} - 10^{15}\,$GeV.
GUT particles would decay to leptons and quarks, producing 
electromagnetic and hadronic
cascades whose eventual tertiary products
would cosmic rays or $\gamma$-rays.
Under certain assumptions, the cosmic
rays 
would be produced close enough to Earth to avoid the
GZK cutoff.

Another interesting possibility is that of primordial black holes.
\index{primordial black holes}
Black holes are not truly black, but emit a spectrum of
radiation \cite{Hawking}.
\index{Hawking radiation}
As the black hole loses energy, it gets hotter, and this results in
more available quantum states for the emission.
Eventually, the black hole evaporates in a final explosion
that releases a burst of radiation.
Small black holes created in the early Universe with
masses near $10^{14}\,$g would be in the process of evaporation
now.
Such primordial black holes would produce bursts of
VHE $\gamma$-rays on time scales
of microseconds to seconds \cite{Zas}.

The search for evidence of supersymmetry is a very 
important frontier area of particle physics.
Given the possibility that supersymmetry may manifest itself at
TeV energies,
it is natural to consider VHE astrophysical signatures.
One possible signature would be the direct detection of
neutral supersymmetric particles.
For example, it has been proposed that the highest energy
cosmic rays are supersymmetric strongly interacting particles
($S^0$) having a much greater path length through
the CMBR than ordinary nucleons \cite{Farrar}.
A second possible signature for supersymmetry would be the
detection
of secondary decay products from a supersymmetric particle.
The most interesting possibility here
is that of the neutralino, 
\~X$^0$.
\index{neutralino-astrophysical detection}
Neutralinos may be
the lightest supersymmetric particles and they may
also comprise a large fraction of the dark matter in the galaxy
\cite{Jungman}.
Since neutralinos must couple to ordinary matter through
weak interactions, the weak scale cross-section determines the
temperature at which the neutralino would freeze out,  
and hence its relic abundance.
Neutralinos concentrating in the galactic center would
annihilate via loop processes,
\~X$^0$\~X$^0\,\rightarrow\,\gamma\gamma\,,\,Z\gamma$,
yielding a
very high energy $\gamma$-ray signature 
\cite{Bergstrom}.
Similarly, neutrinos would be produced from the annihilations of
gravitationally trapped neutralinos in the center of the Sun or
the Earth.
Estimates for the $\gamma$-ray or neutrino fluxes 
are sensitive to parameters of the given model of supersymmetry,
to the mass of the neutralino, and to the density profile of 
galactic dark matter.
In spite of large uncertainties in expected flux levels,
the search for supersymmetry by
these techniques is important for several reasons:
1) the signature,  a highly directional, mono-energetic
flux of $\gamma$-rays or neutrinos, would be unambiguous,
(i.e. it would provide
``smoking gun'' evidence for supersymmetry),
2) a statistically significant detection would allow
measurement of the neutralino mass, and
3) the next generation of $\gamma$-ray and neutrino telescopes 
are sensitive to neutralino masses up to several TeV, which may
be out of reach of accelerator experiments for some time to come.

\section{The Experimental Situation}

\subsection{Detection Techniques: $\gamma$-rays and Cosmic Rays}

\begin{figure}[htb]
\begin{center}
\epsfig{file=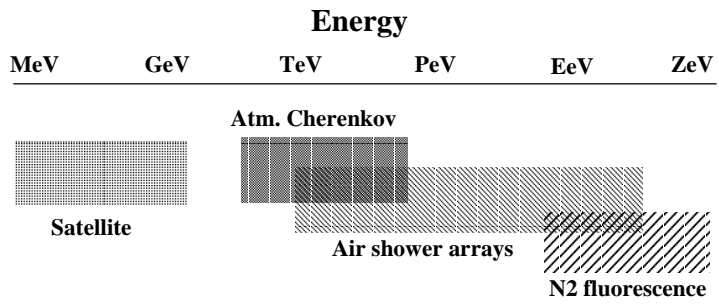,height=1.6in}
\caption{Detection techniques for 
high energy $\gamma$-ray and cosmic ray experiments.
For each technique, an approximate energy range for 
current instruments is shown.
For cosmic rays, the sensitivity of
the satellite detectors extends
to TeV energies.}
\end{center}
\label{technique}
\end{figure}

As shown in Figure~\ref{technique},
$\gamma$-rays and cosmic rays span an enormous range of energies,
from $1\,$MeV to $10^{20}$\,eV.
Given this range, a single detection technique will not suffice.
Satellite and balloon experiments above
the Earth's atmosphere operate at MeV and GeV energies.
At high energies, the particle flux is small enough
so that space-borne instruments become flux limited.
For $\gamma$-rays, the practical upper limit of sensitivity for
the instruments on
the Compton Gamma Ray Observatory is $\sim 20\,$GeV.
For cosmic rays, the upper range of space-borne
detectors is higher by several orders of
magnitude due to the greater cosmic ray flux.

At energies above the reach of balloon and satellite experiments,
we use the Earth's atmosphere as the target and absorber
of high energy particles.
Primary
$\gamma$-rays and cosmic rays interact in the atmosphere
to create extensive air showers that propagate to the ground.
Ground-based detectors sample the
Cherenkov radiation, the charged particles, or the
blue/UV fluorescence from nitrogen excitations
to determine the arrival directions and energies of
the incoming primary particles.
As shown in Figure~\ref{technique},
atmospheric Cherenkov telescopes,
air shower arrays,
\index{atmospheric Cherenkov telescopes}
\index{air shower arrays}
and 
Nitrogen fluorescence detectors
\index{nitrogen fluorescence}
operate in various energy ranges from $200\,$GeV to
$10^{20}\,$eV.

\subsection{$\gamma$-ray Experiments and Results}

There are a number of ground-based $\gamma$-ray telescopes
in operation around the world.
This paper does not attempt to be comprehensive, and
several recent articles have reviewed
the experimental situation \cite{Ong,Hoffman}.
The state-of-the-art Cherenkov telescopes include
Whipple in Arizona, USA,
\index{Whipple Observatory}
HEGRA on the island of La Palma,
\index{HEGRA}
CAT at Themis, France,
\index{CAT}
and CANGAROO at Woomera, Australia.
\index{CANGAROO}
Operating air shower arrays include
HEGRA and the Tibet Array in Yanbajing, Tibet.

\begin{figure}[htb]
\begin{center}
\epsfig{file=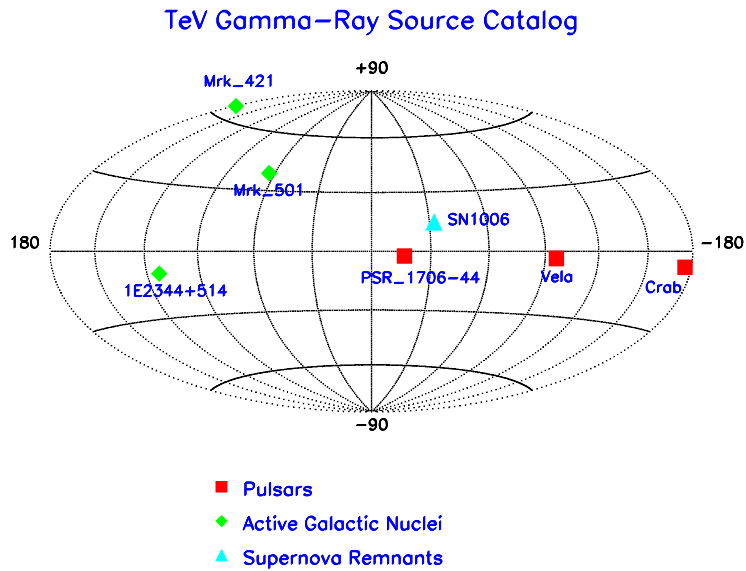,height=3.0in}
\caption{Current source map for the
TeV $\gamma$-ray sky.
The various source classifications are given
by the legend.} 
\end{center}
\label{gammamap}
\end{figure}

$\gamma$-ray astronomy is a quickly changing field with
a variety of sources and new phenomena discovered
during the last decade.
The EGRET experiment 
\index{EGRET}
on the Compton Gamma Ray Observatory
has detected $\sim 150$ sources at energies between
30\,MeV and 20\,GeV.
Many of the EGRET sources remain unidentified, but others have
been associated with pulsars and AGN.
At energies above 200\,GeV, the current status of the field
is shown in Figure~\ref{gammamap}.
Atmospheric Cherenkov telescopes have
detected at least seven VHE sources and 
five additional sources have been tentatively identified.

The sources detected at very high energies can be
categorized as pulsar nebulae, AGN, and SNRs.
The detection of each different source category
was a significant advance which brought us new information
about high energy astrophysical phenomena \cite{Weekes}.
These detections have also raised numerous questions.
For example, for AGN,
remarkable emission has been detected from a few nearby sources,
but many questions remain about how these sources work and
why we cannot see more distant objects.

For the future,
there are a variety of important areas that need exploration:
\begin{enumerate}
\item To date, no sensitive experiments have operated in the
energy range between 20 and 200 GeV.
New instruments, and possibly new techniques, are needed
to explore this energy band where the likelihood
of exciting new astrophysics is high.
For example, it is in this energy range that we
expect to measure spectral features
resulting from the absorption of $\gamma$-rays from
by the cosmic infrared radiation
\cite{Stecker}. 
\item So far, it appears that we have detected
very high energy radiation from the most luminous astrophysical
sources.
Significant increases in flux sensitivity should lead to
the detection of fainter objects and more types of sources.
\item The most important $\gamma$-ray results at energies above
200\,GeV have come from atmospheric Cherenkov telescopes
which have limited fields of view and operate only on dark,
clear nights.
We need instruments which have wide fields of view and operate
with close to 100\% duty cycle.
Such instruments would be able to detect
dramatic transient phenomena, such as gamma ray bursts.
\end{enumerate}

To address the major goals for the future,
there are a number of new high energy $\gamma$-ray telescopes
soon to come on line or being proposed for the future.
To extend the reach of ground-based telescopes to lower
energies, the STACEE and
\index{STACEE}
CELESTE experiments each use 
large
\index{CELESTE}
arrays of solar heliostat mirrors to
detect fainter Cherenkov showers and thus to observe $\gamma$-rays at
lower energies.
Both experiments are in the final stages of construction
and have reported promising initial results \cite{Oser,Smith}.
The MAGIC imaging Cherenkov
\index{MAGIC} 
telescope \cite{Magic}, which consists
of a large 17\,m reflector and a state-of-the-art 
camera, is now under construction.
MAGIC hopes to achieve an effective energy threshold as low
as 20\,GeV.

A new air shower experiment that has recently come on
line is MILAGRO \cite{Sinnis}.
\index{MILAGRO}
Consisting of a large man-made pond of purified water
viewed by $\sim 1000$ photomultiplier tubes,
MILAGRO detects air showers at a median energy threshold
of 1\,TeV.
By using the air shower technique, MILAGRO will carry out the
first all-sky survey at TeV energies.
The encouraging initial results from a prototype of MILAGRO
include the tentative detection of Mrk 501 
and possible correlated TeV $\gamma$-ray emission
from a gamma ray burst \cite{Milagrito1}.

\begin{figure}[htb]
\begin{center}
\epsfig{file=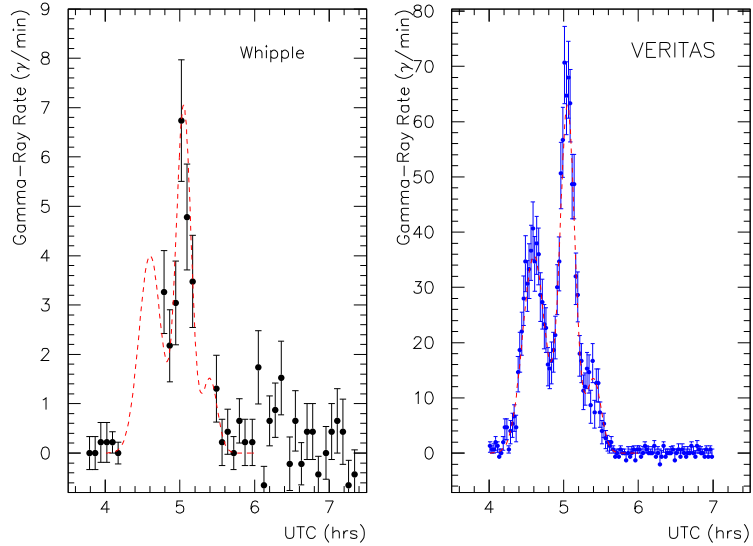,height=2.8in}
\caption{Left: response to a rapid AGN flare in 1996 by
the Whipple Cherenkov telescope
\protect\cite{Whippleflare}.
The dashed curve corresponds to a hypothetical
intrinsic light curve consistent with the
measured data.
Right: simulated response of the VERITAS array
to the same hypothetical light curve.
VERITAS detects a much larger signal with
much finer time resolution than Whipple.}
\end{center}
\label{veritas_flare}
\end{figure}

In order to substantially improve the sensitivity,
and to extend the energy range, of the atmospheric
Cherenkov technique, several groups are developing
new telescopes that employ arrays of imaging reflectors
\cite{Hofmann}/
HESS,
\index{HESS}
initially consisting of four 12\,m reflectors in Namibia,
and Super-CANGAROO,
\index{Super-CANGAROO}
consisting of four 10\,m reflectors located
in Australia, will be the premier ground-based
$\gamma$-ray telescopes in the Southern Hemisphere.
In the Northern Hemisphere, VERITAS
\index{VERITAS}
\cite{Veritas}, is being proposed as an array of seven 10\,m
reflectors.
As shown in Figure~\ref{veritas_flare},
the dramatic improvement in sensitivity of VERITAS
over Whipple
(which represents the current generation of atmospheric
Cherenkov telescopes) will allow for
much more detailed study of rapid very high
energy phenomena.

\begin{figure}[htb]
\begin{center}
\epsfig{file=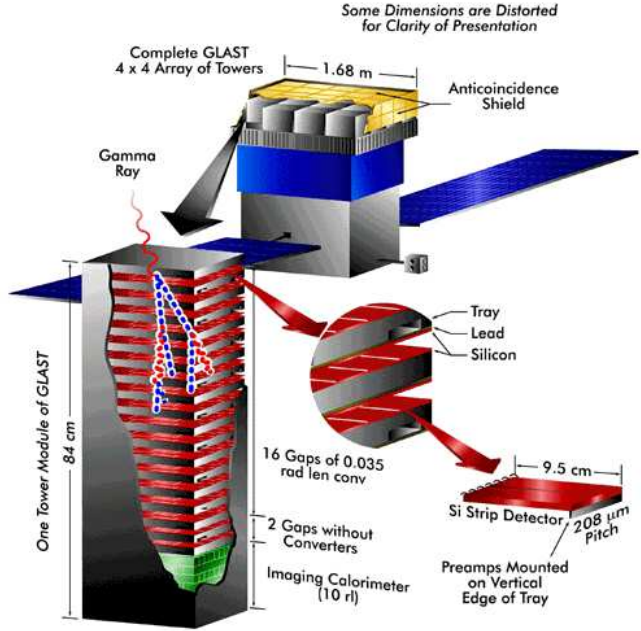,height=3.4in}
\caption{Artist's conception of GLAST, the
next major satellite $\gamma$-ray telescope
\protect\cite{Gehrels}.
The main GLAST instrument will consist of
interleaved Si-strip/absorber layers for tracking
and an imaging CsI calorimeter for energy and position
measurement.}
\end{center}
\label{glast}
\end{figure}

The single most important new $\gamma$-ray telescope
will be flown in space.
GLAST \cite{Gehrels}, will be a state-of-the-art
\index{GLAST}
detector using many techniques of
experimental particle physics, such
as Si-strip tracking and CsI calorimetry.
With a very wide field of view, and a
suitable pointing strategy, GLAST will scan
the entire sky on every orbit, offering
unparalleled coverage of transient $\gamma$-ray
phenomena, such as AGN flares and gamma ray bursts.
GLAST will have substantially improved
characteristics 
(angular resolution, energy resolution, energy range, etc.)
relative to its predecessor, EGRET.
The resulting improvement in sensitivity of GLAST will
enable the detection of up to two orders of magnitude more
sources (e.g. approximately 3000-4000 AGN).
An artist's conception of GLAST is shown in
Figure~\ref{glast}.

\subsection{Cosmic Ray Experiments and Results}

The variety of experimental techniques for the detection of
high energy cosmic rays is summarized in Figure~\ref{technique}.
Satellite instruments have measured the elemental composition
of the cosmic rays at energies up to $10^{14}$\,eV.
An important outstanding issue concerns the level of antimatter,
e.g. positrons, antiprotons, antihelium, etc.
A primordial source of antimatter would be an important
discovery.
Searching for this source is one of the main goals of
the Alpha Magnetic Spectrometer (AMS) satellite detector.
\index{Alpha Magnetic Spectrometer}
In June 1998, AMS took initial cosmic ray data during
a flight on the  space shuttle Discovery. 
Initial results based on
the AMS data have been presented \cite{AMS}.
AMS is now in the process of reconfiguration for an extended
physics program on the International Space Station.

At energies above $10^{14}$\,eV, ground-based air shower detectors
provide the only effective way to carry out cosmic ray measurements.
Determining the cosmic ray composition
by indirect detection techniques has been notoriously difficult.
No clear understanding of the composition exists in the energy region
near the bend in the energy spectrum 
(the ``knee'' at $\sim 3\times 10^{15}\,$eV).
It is in this region, however,
 where we expect to learn something about
the origin of cosmic rays \cite{Watson}.
Future ground-based and satellite experiments may shed light
on this difficult, but important, problem.

For the highest energy cosmic rays above the GZK cutoff,
the current experimental situation is 
summarized in Figure~\ref{agasa}.
There is good evidence that the cosmic ray spectrum extends
to $10^{20}$\,eV and beyond, but the statistics are not
overwhelming.
It is important to keep in mind that the flux of particles
above $10^{20}$\,eV is approximately 1 per km$^2$ per century!
\index{AGASA}
The AGASA experiment, which has detected the majority of the
events beyond the GZK cutoff, has a collection factor
of approximately 120 km$^2\cdot$sr.
The newly commissioned Fly's Eye HiRes experiment 
improves upon this factor by an order of magnitude.
\index{Fly's Eye}
Fly's Eye HiRes consists of two fluorescence detector sites
located 13\,km apart.
Each detector employs spherical
mirrors to reflect air shower fluorescence light onto 
photomultiplier tube cameras.
The construction for the experiment has been completed, and
preliminary results based on data-taking with an initial
configuration have been reported \cite{Jui}.
The early Fly's Eye 
results provide clear confirmation of the AGASA result
that the cosmic ray 
spectrum continues to $10^{20}\,$eV, and beyond.

\begin{figure}[htb]
\begin{center}
\epsfig{file=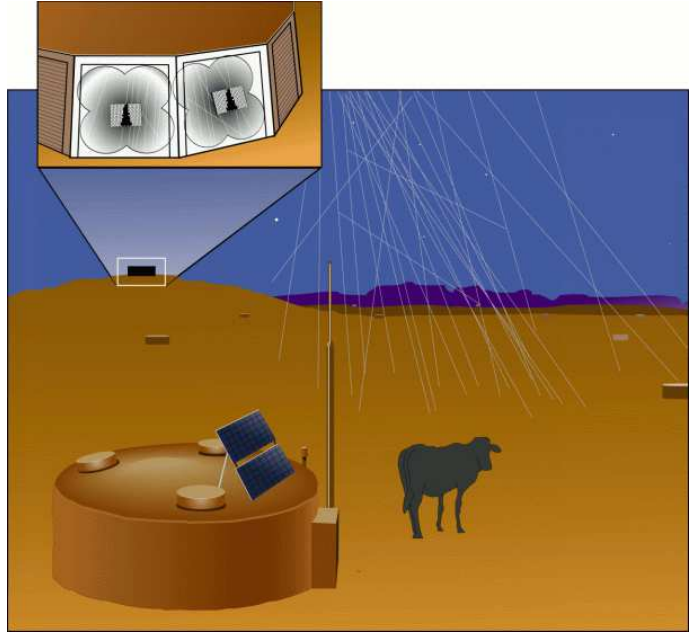,height=3.3in}
\caption{Artist's conception of the Auger Project
for the detection of
the highest energy cosmic rays
\protect\cite{Boratav}.
The water tank in the foreground is one of 1600
such tanks used to detect the air shower particles.
A nitrogen fluorescence detector (one of three) is
shown in the background.
Two Auger sites are evisioned, one in each of the
Northern and Southern Hemispheres.}
\end{center}
\label{auger}
\end{figure}

The exciting results relating to the highest energy cosmic rays
have prompted the development of new, and even larger,
cosmic ray experiments.
The Auger Project plans to construct two giant air shower arrays,
one each in the Southern and Northern Hemispheres.
\index{Auger Project}
The arrays
incorporates both particle and nitrogen fluorescence 
detectors \cite{Boratav}.
Each array consists of 1600 particle detection stations
on a grid covering approximately 3000 km$^2$.
Three fluorescence detectors are used to improve the
energy calibration of the ground array.
As shown in Figure~\ref{auger}, a detector station employs
a large water tank viewed by self-contained instrumentation
to record the arrival of the air shower particles.
The Southern Hemisphere array is currently under construction
in Mendoza, Argentina.

The Northern Hemisphere array of Auger is planned for construction in
Millard County, Utah.
A very large fluorescence detector, the Telescope Array,
is also being considered for construction in the same general
region as Auger.
A possible future satellite experiment,
OWL/Airwatch \cite{Scarsi},
\index{OWL/Airwatch}
would detect giant air showers using a downward-looking
fluorescence detector.
This ambitious instrument is currently under development
by groups in the United States and Italy.

\subsection{Very High Energy Neutrino Astrophysics}

As pointed out by Greisen forty years ago \cite{Greisen},
there is an obvious connection between $\gamma$-ray and
neutrino astronomy.  
Both fields involve the detection of neutral particles
produced as secondaries in high energy astrophysical
accelerators.  
In many cases, the same sources give rise to 
fluxes of high energy $\gamma$-rays and neutrinos.

There are important differences, however, between
photon and neutrino astronomy.
Neutrinos have the great advantage over $\gamma$-rays in
that their very small interaction cross section allows
them to travel unimpeded from their production sites
to Earth.
Thus, in principle, neutrino astronomy allows us
to probe the dense central regions 
of objects such as gamma ray bursts, AGN,
and supernovae.
We can also
search for neutrino signals coming from the annihilation
of dark matter concentrated
at the center of the Earth and Sun.
Neutrino astronomy may also shed light on important
aspects of particle physics.
A clear demonstration of this possibility is
the evidence for the oscillation of neutrinos produced
by cosmic ray interactions in our atmosphere.

The most probable source of high energy astrophysical 
neutrinos is from the decay of charged pions produced
in a hadronic beam.
Thus, unlike $\gamma$-rays, which can arise from electromagnetic
processes such as synchrotron radiation and inverse-Compton
scattering, neutrinos will likely arise from astrophysical
sources containing proton beams.
Although we have not yet conclusively identified
sources such as these, we know from the flux of cosmic rays
that energetic 
hadrons are produced {\em somewhere} in large quantities.

The fact that neutrinos have a very small interaction cross
section also poses a big difficulty for the
experimentalist in that very large detectors are required.
So far, this difficulty has limited the astrophysical
information learned from neutrino telescopes.
It is difficult to predict for certain how large
experiments must be in order to detect neutrinos from
AGN or gamma ray bursts.
The appropriate detector collection area is probably
1\,km$^2$, or perhaps even larger \cite{Stanev}.

\begin{figure}[htb]
\begin{center}
\epsfig{file=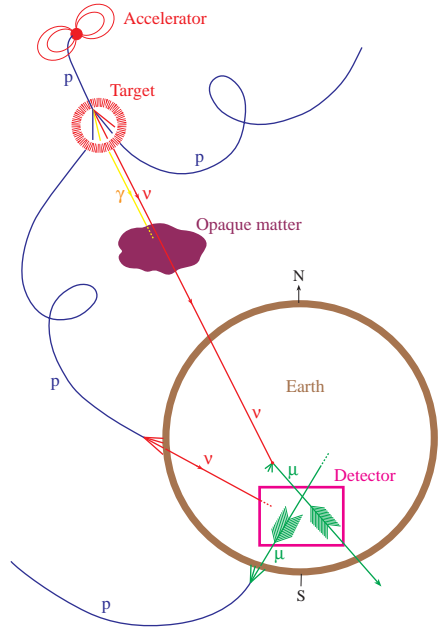,height=3.7in}
\caption{Detection technique for high energy neutrino
telescopes \protect\cite{Halzen2}.
Astrophysical or atmospheric neutrinos penetrate the
Earth to produce upward going muons in the telescope.
The telescope consists of strings of photomultiplier
tubes that detect the Cherenkov radiation produced by
the relativistic muon.
The detector medium could be ice, as in the case of
AMANDA, or deep sea water, as in the case of ANTARES.}
\end{center}
\label{amanda}
\end{figure}

For some time, it has been recognized that clear water
(liquid or ice) would make a very suitable medium for
neutrino detection.
The basic idea is to instrument a large volume
of water with photomultiplier tubes arranged on
long strings.
Upward-going neutrinos (coming from astrophysical sources
on the other side of the Earth) interact in the large volume
of Earth below the detector.
The high energy muons produced from these interactions
are detected via their Cherenkov radiation
in the instrumented volume.
The general method is shown in Figure~\ref{amanda}.
The upward-going signature is required to reduce the
very large ($> 10^5$) background of downward-going muons
produced in cosmic ray air showers.
The cosmic rays also produce a flux of atmospheric neutrinos on
the other side of the globe that can be used as a
calibration source for the detector.

\begin{figure}[htb]
\begin{center}
\epsfig{file=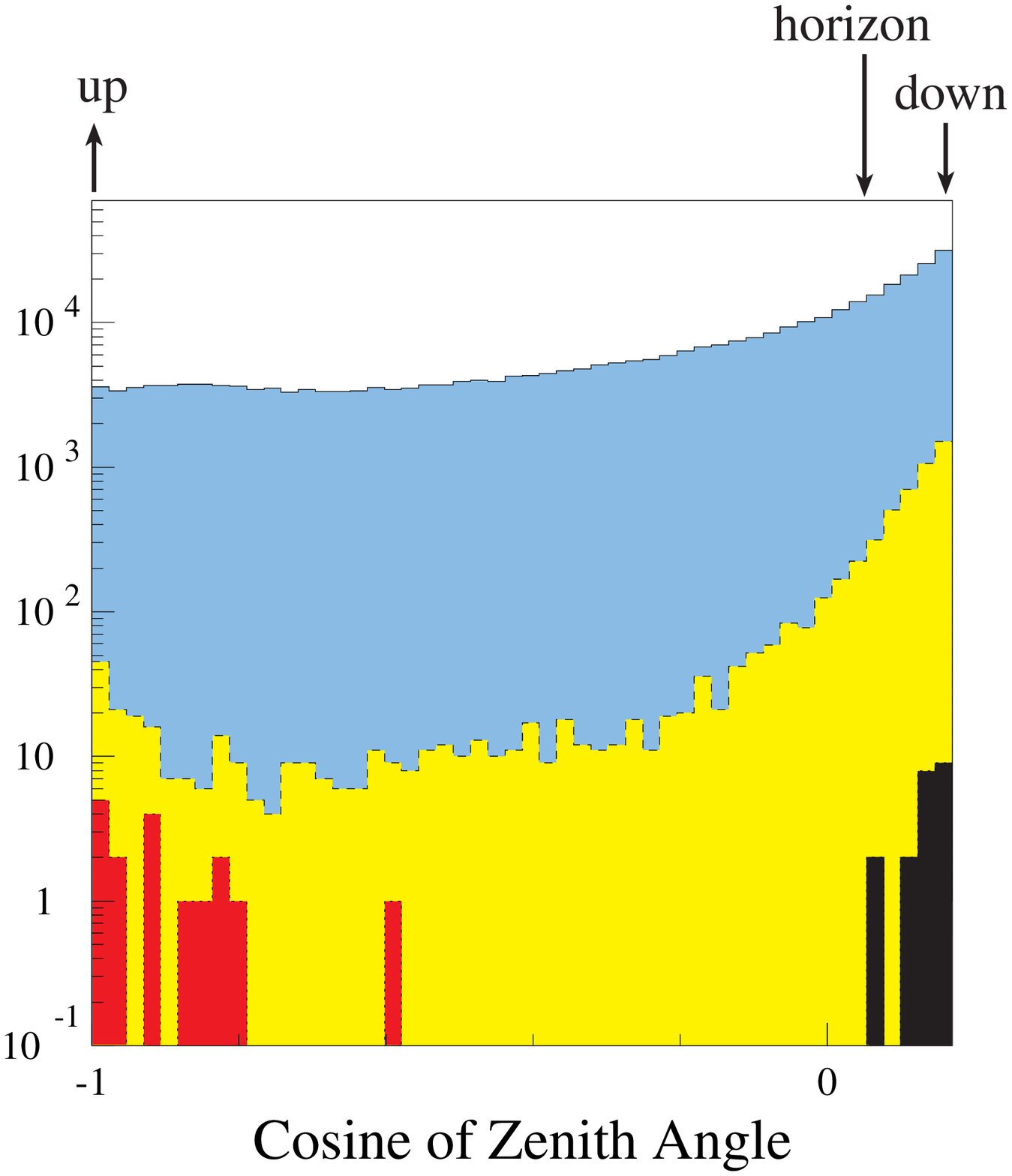,height=3.3in}
\caption{Zenith angle distribution for events
detected by the AMANDA-B telescope \protect\cite{Andres1}.
The three different shadings indicate the distribution
after successive quality cuts are applied to the data.
The events in the dark shaded histogram
near $\cos(\theta) = -1$
correspond to 17 upward-going muon tracks produced
by atmospheric neutrinos.}
\end{center}
\label{zenith}
\end{figure}

Early pioneering work in building very large neutrino telescopes
was carried out by the DUMAND and BAIKAL collaborations.
Indeed, BAIKAL was the first detector to observe
\index{BAIKAL} 
very high energy atmospheric neutrinos \cite{Baikal}.
Presently, the AMANDA installation at the South Pole represents
the current state-of-the-art in high energy neutrino 
telescopes.
\index{AMANDA}
AMANDA has grown in stages during the last five years.
The current instrument, AMANDA-II,
comprises 675 optical modules 
on 19 strings located at ice depths of 1450 to 2050 meters.
AMANDA-II
has a collection area of approximately 30,000 m$^2$ 
(\@ 1 TeV).
An earlier version of the detector, AMANDA-B, 
operated between 1997 and 1999 and 
recorded a large data set.
As shown in Figure~\ref{zenith},
by means of stringent quality cuts on the track fitting,
a clean sample of 17 muon tracks due to
atmospheric neutrino events has
been isolated \cite{Andres1}.
This result demonstrates that AMANDA can detect a neutrino
signal.

To increase the likelihood of detecting astrophysical neutrinos,
several groups have proposed water Cherenkov telescopes on much
larger scales than AMANDA.
There are three projects being considered for deployment
in the deep Mediterranean water:
ANTARES\cite{Antares}, NESTOR\cite{Nestor}, and NEMO.
\index{NESTOR}
Of these various efforts, the French
\index{ANTARES}
ANTARES project is perhaps the furthest along.
The ANTARES group has
secured approval to build a telescope consisting of
1000 optical modules arranged on 13 vertical strings.
The modules will deployed at water
depths of 2100 to 2400\,m, 30\,km off shore.
The baseline design for ANTARES calls for 
an effective collection area of approximately
0.1\,km$^2$. 
The neutrino energy threshold
should be below 100\,GeV which will permit the
study of atmospheric neutrino oscillations, 
in addition to the search for neutrinos from astrophysical sources.

A very large km$^3$ size detector is being considered
for deployment in the South Pole ice.
The IceCube collaboration
\index{IceCube}
is proposing an array of 81 detector strings arranged on
a square grid of 125\,m string spacing \cite{IceCube}.
Each string would hold 60 optical modules with a vertical
module spacing of 16\,m.
The effective neutrino energy for IceCube of approximately
400\,GeV would
be somewhat higher than for ANTARES, but IceCube will
have a very large collection area ($1\,$km$^2$ \@ 10 TeV) for
neutrinos from astrophysical sources.

\section{Summary}

Astronomy using very high energy particles ($\gamma$-rays, cosmic
rays, and neutrinos) is a diverse and rapidly developing
field.
It is currently a field largely driven by experimental results
where significant scientific progress
can be made on a relatively short time scale.
For example, in
the last few years there have been several exciting discoveries.
Using $\gamma$-rays, we are probing remarkable and unexpected phenomena
in objects such as active galaxies and gamma ray bursts.
We are also searching for the origins of the cosmic radiation.
At the very highest energies, we
are discovering cosmic ray particles that probably should not
be there, but are.
Discoveries in this field often raise as many questions as they answer.

For the future, there will be an expanding interest in this field,
both to understand astrophysics under extreme conditions and to
search for evidence of physics beyond the standard models of
elementary particles and cosmology.
Future next-generation experiments in space and on the ground will
greatly expand our discovery horizon.
For $\gamma$-rays, the new projects include more powerful
atmospheric Cherenkov telescopes and a new $\gamma$-ray satellite,
GLAST.
Larger air shower detectors, such as Fly's Eye HiRes and Auger,
will explore questions relating to the highest energy cosmic rays.
In the world of neutrinos, the AMANDA experiment has 
demonstrated the ability to detect neutrinos produced in the
atmosphere.
Future experiments, such as IceCube and ANTARES, will greatly expand the
possibility of detecting high energy
astrophysical neutrino sources.
We can only hope that our overall knowledge of high energy
particles from the Universe advances at the same rate
as new instruments are being constructed.

\bigskip
I wish to acknowledge the help and encouragement of many people
in the particle physics and astrophysics communities.
The contributions of the following people were particularly
important:
Katsushi Arisaka,
Steve Barwick,
Michael Catanese,
Corbin Covault,
Jim Cronin,
Francis Halzen,
Charles Jui,
Tadashi Kifune,
Peter Leonard,
Eckart Lorenz,
John Matthews,
Masaki Mori,
Gus Sinnis,
David Smith,
Simon Swordy,
Steve Ritz,
Pierre Sokolsky,
Masahiro Takeda,
Masahiro Teshima,
Trevor Weekes,
and Heinz V\"olk.
I also thank the organizers of the Lepton-Photon Symposium
(especially John Jaros, Helen Quinn, and Michael Peskin)
for their encouragement and patience.
Any inaccuracies are my fault alone.
This research is supported in part by the
National Science Foundation.

\end{document}